\documentclass[12pt]{article}
\usepackage{amsmath}
\usepackage{graphicx}
\usepackage{enumerate}
\usepackage{natbib}
\usepackage{url} 
\usepackage{longtable}
\usepackage{comment}
\RequirePackage{amsthm,amsfonts,amssymb}
\RequirePackage[colorlinks,citecolor=blue,urlcolor=blue]{hyperref}
\RequirePackage[ruled,linesnumbered]{algorithm2e}
\newcommand{\blind}{1}

\begin{document}

\newtheorem{theorem}{Theorem}[section]
\newtheorem{proposition}[theorem]{Proposition}
\newtheorem{definition}[theorem]{Definition}
\newtheorem{rema}{Remark}

\allowdisplaybreaks[4]
\def\spacingset#1{\renewcommand{\baselinestretch}%
{#1}\small\normalsize} \spacingset{1}
\if1\blind
{
  \title{\bf Multiple change point detection in tensors}
  \author{Jiaqi Huang\thanks{
    The research was supported by a grant from the National Natural Foundation of China.
Jiaqi Huang is the first author, and Lixing Zhu is the corresponding author.
}\hspace{.2cm}\\
    School of Statistics,  Beijing Normal University\\
    and \\
    Junhui Wang \\
    Department of Statistics, Chinese University of Hong Kong\\
    Lixing Zhu\\
    School of Statistics, Beijing Normal University\\
    Xuehu Zhu\\
    School of Mathematics and Statistics, Xi'an Jiaotong University}
  \maketitle
} \fi
\if0\blind
{
  \bigskip
  \bigskip
  \bigskip
  \begin{center}
    {\LARGE\bf Multiple change point detection in tensors}
\end{center}
  \medskip
} \fi

\bigskip

\begin{abstract}
This paper proposes a criterion for detecting change structures in tensor data. To accommodate tensor structure with structural mode that is not suitable to be equally treated and summarized in a distance to measure the difference between any two adjacent tensors, we define a mode-based signal-screening  Frobenius distance for the moving sums of slices of tensor data to handle both dense and sparse model structures of the tensors. As a general distance, it can also deal with the case without structural mode. Based on the distance, we then construct  signal statistics using the ratios with adaptive-to-change ridge functions.  The  number of changes and their  locations can then be consistently estimated in certain senses, and the confidence intervals of the locations of change points are constructed. The results hold when the size of the tensor and the number of change points diverge at certain rates, respectively. Numerical studies are conducted to examine the finite sample performances of the proposed method. We also analyze two real data examples for illustration.
\end{abstract}
\noindent%
{\it Keywords:}  Adaptive-to-change ridge; Change structure of tensor; Ridge ratio criteria; Signal-screening  Frobenius distance; Mode-based signal-screening  Frobenius distance
\vfill

\newpage
\spacingset{1.9} 
\section{Introduction}
\label{sec:intro}

%

 One of the challenges in the age of big data is heterogeneity. The complexity of the data generation mechanism cannot be fully captured by classical statistical models aimed at studying independent and identically distributed data. Significantly when the data are collected over time, the generation mechanism may change over time, leading to possible structural changes. A commonly adopted assumption is that data structure may only change at times while remaining stationary between two adjacent change points.

   This paper focuses on tensor data.
  Different definitions of structural changes  adapt to the nature of applications. For example, \cite{7452353}  identified significant network structure changes with all subjects over time by using
 the Grassmann distance between the low-rank approximations of two tensors at adjacent times. 
   Their method assumes the  normality on the data distribution. 
 \cite{al2017tensor} constructed two tensors through adjacent matrices and used a Frobenius norm-based normalized cost function to identify change points. 
\cite{zhan2021tensor} performed HSI-MSI fusion to obtain high-resolution spectral and spatial images, calculated the differences between the fusion images at two different time points, and used the classification method. Based on the Frobenius distance, the generalized tensor regression method  divides the differences into several small tensors.
      Though empirical successes have been reported,
      formal theoretical guarantees of the above methods  still need to be improved. 

For dynamic network data 
or order-two tensors, \cite{wang2021optimal} and \cite{zhao2019change} respectively defined the CUSUM-based and MOSUM-based Frobenius distances and chose respective thresholds to identify local maximizers associated with locations of the change points.
 Both methods need to deal with consistently estimating the sparsity level and choosing thresholds to handle the unknown magnitudes of changes.
Specifically, to identify all possible local maximizers associated with the locations of the change points, the thresholds in the MOSUM-based methods proposed by \cite{zhao2019change} and  \cite{wang2021optimal} are related to the unknown minimum magnitude of the changes. Thus, they gave two recommendations for determining the threshold  in practice. 


 We note that in the literature, the Frobenius norm has been popularly used to measure the distance between adjacent data. This distance  over all elements of a tensor, including vector and matrix, can successfully summarize the  information on changes. On the other hand, various research works have also pointed out that some mode of tensor may represent specific structural information and should be treated separately. The examples include the sensor mode in the seismic Data set of earthquake forecasting (
\cite{xie2019asynchronous},  \cite{chen2022high}), 
the subject mode in the FMRI data set (\cite{dai2019discovering}), the sub-region mode in the Brain cancer incidence data set about New Mexico  (\cite{braincancer}), and the spatial grid mode in the  data set of the 50-year high-resolution monthly gridded precipitation  over Spain (\cite{herrera2012development}). The research interests  focus on the significant structural changes with all sensors,  subjects/individuals, sub-regions and spatial grids  over time
respectively. It is thus inappropriate to treat these structural modes equally as other modes in the tensors, as to be summarized in the existing Frobenius distance. To fully leverage the
information on the structural mode, our idea in this paper is to first define a new Frobenius-type distance for multiple sub-tensors along the mode, and then take the minimum over all
the defined Frobenius distances as the final criterion to detect the heterogeneity of tensors.
To achieve this goal, we propose a mode-based signal-screening Frobenius distance over the sub-tensors in this paper.  When  tensors have no structural mode or there is no need of separately handling structural modes, this  general distance can be naturally defined over all elements. Our contributions can be summarized as follows.

   First, to adapt dense and sparse model structures automatically, \cite{wang2021optimal}, for the dynamic networks,  introduced a notion of sparsity level, but how to consistently estimate this level remains unsolved. We, in this paper, define a novel mode-based signal-screening  Frobenius distance between two adjacent tensors in a sequence of the moving sums (MOSUM) as the measure for changes.  Fixing the structural mode, the new distance is the averaged  Frobenius distance over the remaining signal elements after ruling out those non-signal elements of `small' values of MOSUM (see, e.g., \cite{cho2015multiple}).  The resulting distance becomes a vector.
    We write it as the MSF-distance in short. This distance can be guarded against being spuriously large, which may cause overestimation when the number of elements is large (see, e.g., \cite{cho2015multiple}), or spuriously small, which may cause underestimation if the average over its number of elements is considered.  
 As a generic methodology,  it can  be applied to tensors of order-one (vector), order-two (matrix), and higher-order. Without signal screening in the distance, the distance is then defined over all elements of tensors.  

%
Second,   most existing MOSUM-based or CUSUM-based methods  rely on the fact that the changes arrive at the maxima and
their estimation consistency can be derived when the threshold is properly determined. As the magnitudes of the changes are usually unknown and the threshold depends on the distribution of  statistic,  its determination is a concern. 
To circumvent this concern, we define a sequence of ratios between any two adjacent MSF-distances with an adaptive-to-change ridge function. As the resulting sequence is a vector each component having the minima of zero arrived at $z_k-2\alpha_n+1$, the maxima of infinity arrived at $z_k-\alpha_n$.
   Here $z_k$'s are the locations of the changes, and $\alpha_n$ is the window size of the moving sums. We then define a signal statistic as the minimum of all the components in the vector sequence.   The choice of thresholding values can be much more flexible (in theory, any value between 0 and 1 when we search for the minima). The computation can be fast as all change points can be detected simultaneously. The plot of  the sequence shows the periodic pattern and can also assist the detection. The estimation consistency for the locations and the number of change points can be derived.  Further,  
 we apply a similar idea as 
in \cite{chen2021inference} to construct the confidence intervals of the locations of change points and give more discussion on its construction in Section~3. 

Third, in the ratios, we use the adaptive-to-change ridge function to enhance the detection capacity of the criterion compared with a constant ridge as in \cite{zhao2020detecting} for detecting univariate mean changes. The original purpose of using diminishing ridge is to make the undefined $0/0$ ratios be $1$ such that the detection becomes easy. The adaptive-to-signal ridge function can then go to infinity in the segments without involving change points. Thus, the values of the signal statistic (the empirical version of the sequence mentioned above) can tend to one at a  much faster rate than that with a constant ridge so that the curve oscillation at the sample level can be alleviated and false changes could be reduced in practice.  

Fourth, in the case with no structural mode and  the distance is then naturally defined over all elements of the tensor, the theoretical properties of the criterion will be different.
 We then separately investigate the  distances  and the corresponding statistics in two scenarios so that we have two signal statistics. 

Fifth, the asymptotic studies on the signal statistic are of theoretical interest. Particularly, when the number of elements in the tensor is not very large, the independent error terms are not required identically distributed, and the second moments are not necessarily identical. These conditions are  mild. 
Although the statistics are empirical versions of their corresponding functions at the population level, inconsistency in some non-negligible segments occurs. Thus, we will give a detailed analysis to identify these segments and study the asymptotic behaviors of the signal statistics in these segments.  We will also put an algorithm in Supplementary Materials, which is to prevent the detected change points from falling into these segments so that the estimated number and locations of changes are still consistent.
Furthermore, the consistency can be established when the required spacing between any two adjacent change points is shorter than in \cite{zhao2020detecting} that  also used ridge ratios to construct a criterion for mean change detection. We also make some comparison  with other methods in the literature for the minimum signal strength and estimation accuracy to see the pros and cons of our method.

   The rest of the paper is organized as follows. Section~2 introduces the problem and notion of change point detection in tensors and suggests a criterion to detect change points. As the motivation of the method, we first consider the distance over all elements.   Section~3 presents the consistency of the estimated number of changes and the estimated locations in a certain sense.  Section~4 includes simulation studies, including the cases with tensors of order-one and higher-order.  We find that our method can have better performance in the detection when the number of tensor elements gets larger. This dimensionality blessing phenomenon, particularly in dense cases, might suggest that the dimensionality might not seriously affect the new methods. 
  Section~5 includes the illustrative analyses for two real data examples. Section~6 discusses the merits and limitations of the new method. 
  All other technical proofs  and some of numerical studies are included in Supplementary Materials.
\section{Methodology Development}
\subsection{Preliminary}
 Let $\mathcal{X}_i\in \mathbb{R}^{p_1\times \cdots\times p_{\kappa}},i=1,2,\cdots,n$ be independent order-$\kappa$ tensors satisfying the model as follows
    \begin{align}\label{model}
       \mathcal{X}_{i}&=E(\mathcal{X}_i)+\mathcal{E}_i,i=1,2,\cdots,n,
    \end{align}
    where $\mathcal{E}_i$'s are independent error terms with zero mean. Notice that we do not require $\mathcal{E}_i$ to be identically distributed or from some specific distributions. {Here, the order $\kappa$ is the number of dimensions, and each dimension is called a mode. The element $(i_1,i_2,\cdots i_{\kappa})$ of tensor $\mathcal{X}_i$ is denoted by $\mathcal{X}_{i,i_1i_2\cdots i_{\kappa}}$. If we fix the mode-$l$ index of the tensor to be $i_l$, then the other $(\kappa-1)$-dimensional sections of a tensor are called the $i_l$-th mode-$l$ slice. 
    Thus, a $\kappa$th-order tensor is a $\kappa$-dimensional array with $\kappa$ modes(see more details in  \cite{bi2021tensors}, \cite{kolda2009tensor} and \cite{venetsanopoulos2013fundamentals}).
    } 

    Assume that there are $K$ change points; that is, $1=z_0<z_1<\cdots z_k<\cdots <z_K<z_{K+1}=n$, which divide the original sequence into $K+1$ segments so that
    $$E(\mathcal{X}_i)=\mathcal{M}^{(k)},i=z_{k-1}+1,\cdots,z_{k}.$$
 Write $\alpha_n^{\star}=\min_{1\leq k\leq K+1}|z_{k}-z_{k-1}|$ as the minimum distance between any two change points.
From (\ref{model}), we can see that the classical change point detection for means (order-one tensor) and covariance matrices (order-two tensor) could be regarded as special cases.

 As mentioned in the introduction, 
our idea is 
to define a sequence of distances along the structural mode, which eventually form the proposed signal
a statistic. But to motivate our method well, we first consider the simple case with no structural mode. Therefore, all modes can be treated equally  to define  the distance  over all elements of the tensor in  subsection~2.2 below. The mode-based distance in the general case will be described in subsection~2.3.
\subsection{Signal screening distance and  a signal statistic}

To construct this distance and a signal statistic, we have the following steps. \\

    {\itshape{The moving sums(MOSUM).}} Define the MOSUM of the tensor sequence as follows:
     \begin{equation}\label{musump}
        \mathcal{D}(i)=\frac{1}{\alpha_n}\left(\sum_{j=i}^{i+\alpha_n-1}E(\mathcal{X}_j)-\sum_{j=i+\alpha_n}^{i+2\alpha_n-1}E(\mathcal{X}_j)\right).
     \end{equation}

As commented before, we  define the following signal-screening metric with a threshold $l_n(s) \to 0$ as $n\to \infty$:
    \begin{equation}\label{ada}
      \left\vert\left\vert \mathcal{A}
      \right\vert\right\vert_s^2=\frac{\sum_{i_1=1}^{p_1}\sum_{i_2=2}^{p_2}\cdots\sum_{i_{\kappa}=1}^{p_{\kappa}}a_{i_1 i_2\cdots i_{\kappa}}^2I(a_{i_1 i_2 \cdots i_{\kappa}}^2>l_n(s))}{\sum_{i_1=1}^{p_1}\sum_{i_2=2}^{p_2}\cdots\sum_{i_{\kappa}=1}^{p_{\kappa}}I(a_{i_1 i_2 \cdots i_{\kappa}}^2>l_n(s))+1/n},
    \end{equation}
    where $a_{i_1 i_2\cdots i_{\kappa}}$ denotes $(i_1,i_2,\cdots,i_{\kappa})$ element of the $\kappa$-order tensor $\mathcal{A}$ and the value $1/n$ in the denominator is to avoid the undefined $0/0$ ratio. Actually the denominator is to approximate the sparsity of a tensor. 
 When $l_n(s)=0$, write $||\mathcal{A}||_{s}^2$ as $||\mathcal{A}||_{0s}^2$.


 In this special case, we call $\| {\cal D}(i) \|_{0s}^2$ as the signal screening distance (SF-distance). Some elementary calculations yield the following properties of $||\mathcal{D}(i)||_{0s}^2$: for each $1\le k \le K$
  $$
  ||\mathcal{D}(i)||_{0s}^2=
  \left\{
  \begin{array}{lll}
  0 &  & z_{k-1}\leq i<z_k-2\alpha_n\\
  \frac{\sum_{l=1}^{\kappa}\sum_{i_l=1}^{p_l}\mathcal{D}_{i_1 i_2\cdots i_{\kappa}}^2(i)}{\sum_{l=1}^{\kappa}\sum_{i_l=1}^{p_l}I\left(\mathcal{D}_{i_1 i_2 \cdots i_{\kappa}}^2(i)>0\right)+1/n} \nearrow  &  &z_{k}-2\alpha_n\leq i<z_{k}-\alpha_n\\
  ||\mathcal{M}^{(k+1)}-\mathcal{M}^{(k)}||_{0s}^2& & i=z_{k}-\alpha_n\\
   \frac{\sum_{l=1}^{\kappa}\sum_{i_l=1}^{p_l}\mathcal{D}_{i_1 i_2\cdots i_{\kappa}}^2(i)}{\sum_{l=1}^{\kappa}\sum_{i_l=1}^{p_l}I\left(\mathcal{D}_{i_1 i_2 \cdots i_{\kappa}}^2(i)>0\right)+1/n} \searrow &  & z_k-\alpha_n< i<z_k,\\
  \end{array}
  \right.
  $$
{ where $\searrow$ and $\nearrow$ are strictly increasing and decreasing, respectively.}
When there is no change point, $||\mathcal{D}(i)||_{0s}^2=0$. When a change comes up, the values of $||\mathcal{D}(i)||_{0s}^2$ fluctuate with a length of $2\alpha_n$, monotonically increases from $z_k-2\alpha_n$ to $z_k-\alpha_n$, taking the maximum at the point $z_k-\alpha_n$ and then  monotonically decreases from $z_k-\alpha_n$ to $z_k$.
\\

\begin{figure}
\begin{center}
 \includegraphics[width=0.7\textwidth]{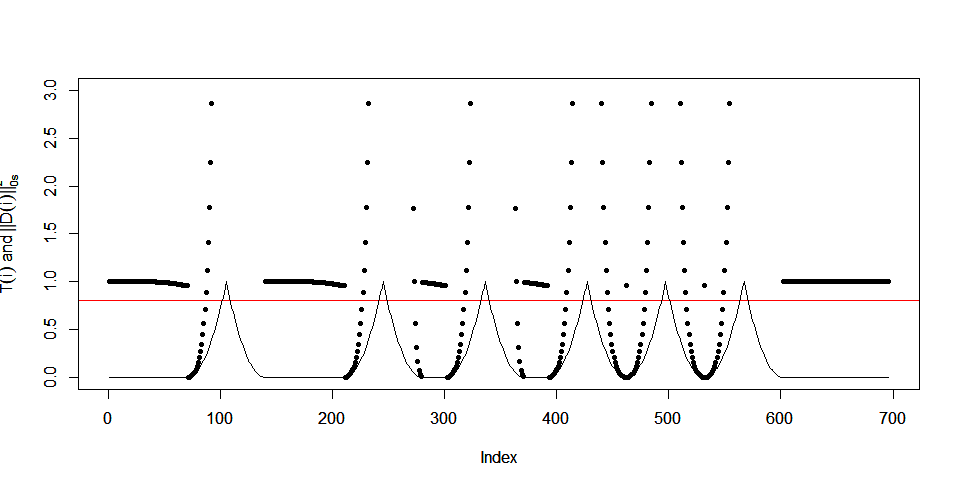}
\end{center}
  \caption{The solid line is for $||\mathcal{D}(i)||_{0s}^2$ and the dotted line for $T(i)$. }\label{population level}
\end{figure}

    {\itshape{Ridge-ratio function.}} According to the properties of $||\mathcal{D}(i)||_{0s}^2$ and $||\mathcal{D}(i+\alpha_n)||_{0s}^2$, we define the ridge ratio statistic at the population level as follows:
    \begin{equation}\label{ratiop}
      T(i)=\frac{||\mathcal{D}(i)||_{0s}^2+c_n^{\star}(i)}{||\mathcal{D}(i+\alpha_n)||_{0s}^2+c_n^{\star}(i)},
    \end{equation}
  where the ridge function $c_n^{\star}(i)$ to be selected later is to  avoid the undefined $0/0$ of  $T(i)$ such that $T(i)$ is close to $1$ in the segments where the moving sums in both $||\mathcal{D}(i)||_{0s}^2$ and $||\mathcal{D}(i+\alpha_n)||_{0s}^2$  involve no locations of change points.
   When  we use an adaptive-to-change ridge function $c_n^{\star}(\cdot)$, Figure~\ref{population level} presents the curves of $||\mathcal{D}(\cdot)||_{0s}^2$ and $T(\cdot)$  for a toy example, showing two very interesting patterns. From the curve of $||\mathcal{D}(\cdot)||_{0s}^2$, we can see that each spike uniquely corresponds to a change point. As discussed in  \cite{zhao2019change} although $||\mathcal{D}(\cdot)||_{0s}^2$ can  indicate the locations and number of changes, the unknown magnitudes of the maxima of $||\mathcal{D}(\cdot)||_{0s}^2$  brings up the issue of  determining a proper threshold.  In contrast, the curve of $T(\cdot)$  with local minima of zero and local maxima of infinity can help determine the threshold.
  Further, 
   different curve patterns of $T(i)$ may occur when the spacings between two adjacent change points are 
   larger than $3\alpha_n$,  $2\alpha_n+f$ or $2\alpha_n$, where $f\in (0,\alpha_n)$. Figure~\ref{population level} gives the basic idea.  More details will be presented in the next section, and the detailed calculation can be found in Supplementary Materials. Briefly speaking, when the spacing is longer than $3\alpha_n$, $T(i)$ monotonically decreases on the left-hand side of $z_k-2\alpha_n+1$, drops to 0 at $z_k - 2 \alpha_n +1$, and increases up to infinity. In the last two scenarios, $T(i)$ monotonically decreases to a local minimum first and then suddenly jumps up to one in the interval with length $f+1,0\leq f <\alpha_n$, and immediately drops down to another local minimum and follows an up to infinity.
 All the local minimizers $z_k's-2\alpha_n+1$ at the right-hand sides of the discontinuous locations in all such intervals correspond to the change points.  As the local minima are zero, the threshold for change point detection can be easily determined.\\

{\itshape{The adaptive-to-change ridge function.}}  Here, we give the formula of the ridge function for a $\nu>1/2$ and any positive constant $s_1$,
\begin{equation}\label{ridge_po}
  c_n^{\star}(i)=\frac{s_1\epsilon_n(\log n)^{\nu}}{I(i\in \mathcal{S}^{\star})+\frac{1}{n}}
\end{equation}
where $\mathcal{S}^{\star}=\left\{i:\sum_{l=1}^{\kappa}\sum_{i_l=1}^{p_l}I([\mathcal{D}_{i_1i_2\cdots i_{\kappa}}(i))]^2>0)>0\right\}$ and $\epsilon_n=\frac{(\log n)^{\frac{1}{2}+\epsilon}}{\sqrt{\alpha_n}}$ that is the uniform rate of $|\mathcal{D}_{n, i_1, \cdots, i_{\kappa}}(i)-\mathcal{D}_{ i_1, \cdots, i_{\kappa}}(i)|$ over all $\{i_1, \cdots, i_{\kappa}\}$. The results will be presented in Section~3. 
First, in the segments involving no locations of changes, the denominator is very small ($1/n$ at the population level); in other words, the ridge tends to infinity such that $T(i)$ tends to one at a fast rate.
Second, in the segments involving the locations of changes,  $c_n^{\star}(i)$ goes to zero with the numerator going to zero and the denominator being larger than $1$.
Further,  $\mathcal{D}(i+\alpha_n)$ around the location $z_k-2\alpha_n$ can be large and $\mathcal{D}(i)$ around $z_k-2\alpha_n$ can still be small.  These lead to small $T(i)$'s around  $z_k-2\alpha_n+1$ such that we can efficiently identify the locations of the changes.  We can also use large $T(i)$'s around  $z_k-\alpha_n$ to identify the locations of the change points. But this is similar to using small $1/T(i)$'s. Thus, in this paper, we do not discuss its use.

%

From Figure~\ref{population level}, we can see that $K$ change points correspond to $K$ disjoint intervals. Thus,  we can determine $K$ disjoint intervals by using a thresholding value for $T(\cdot)$.  Let $(\tilde{m}_k(\tau),\tilde{M}_k(\tau))$ denote the intervals  where $\tilde{m}_k(\tau)$ and $\tilde{M}_k(\tau)$ satisfy the following conditions: for a thresholding value $\tau\in (0,1)$,
 $T(\tilde{M}_k(\tau))<\tau \text{ and }T(\tilde{M}_k(\tau)+1)\geq \tau,$
 and $\tilde{m}_k(\tau)=\tilde{M}_k(\tau)-\frac{2\sqrt{\tau}}{\sqrt{\tau}+1}\alpha_n.$ 

Use the sample versions of $\mathcal{D}(i)$ and $T(i)$ to define the corresponding signal statistic.
Let
    $\mathcal{D}_{n}(i)=\frac{1}{\alpha_n}\left(\sum_{j=i}^{i+\alpha_n-1}\mathcal{X}_j-\sum_{j=i+\alpha_n}^{i+2\alpha_n-1}\mathcal{X}_j\right)$
  and the SF distance-based
  signal statistic is defined as $T_n(i)=\left(||\mathcal{D}_n(i)||_s^2+c_n(i)\right)/\left(||\mathcal{D}_n(i+\alpha_n)||_s^2+c_n(i)\right),$
where $ c_n(i)$ is an estimator of $ c_n^{\star}(i)$: for a $\nu>1/2$ and any positive constant $s_1$,
\begin{equation}\label{cn}
    c_n(i)=\frac{s_1\epsilon_n(\log n)^{\nu}}{I(i\in \mathcal{S})+\frac{1}{n}},
\end{equation}
and  {$\mathcal{S}=\left\{i:\sum_{l=1}^{\kappa}\sum_{i_l=1}^{p_l}I\left(\mathcal{D}_{n,i_1i_2\cdots i_{\kappa}}^2(i)>l_n(s)\right)> 0\right\}$ 
is a replacement of $\mathcal{S}^{\star}$ in $c_n^{\star}(\cdot)$ where   $l_n(s)=s\epsilon_n(\log n)^{1/2}$, and $s$ and $s_1$ are two tuning parameters.}  Specifically, the thresholding value $0$ in $ c_n^{\star}(i)$ is replaced by $l_n(s)$ such that the set ${\mathcal S}^{\star}$ is replaced by $\mathcal{S}$  and $\mathcal{D}(i)$ by $\mathcal{D}_n(i)$.
Note that the results in Section 3 hold for any $\nu >1/2$, but we recommend $\nu=0.55$, which yields satisfactory performance in all the numerical experiments in Section 4.

\begin{figure}
\begin{center}
   \includegraphics[width=0.7\textwidth]{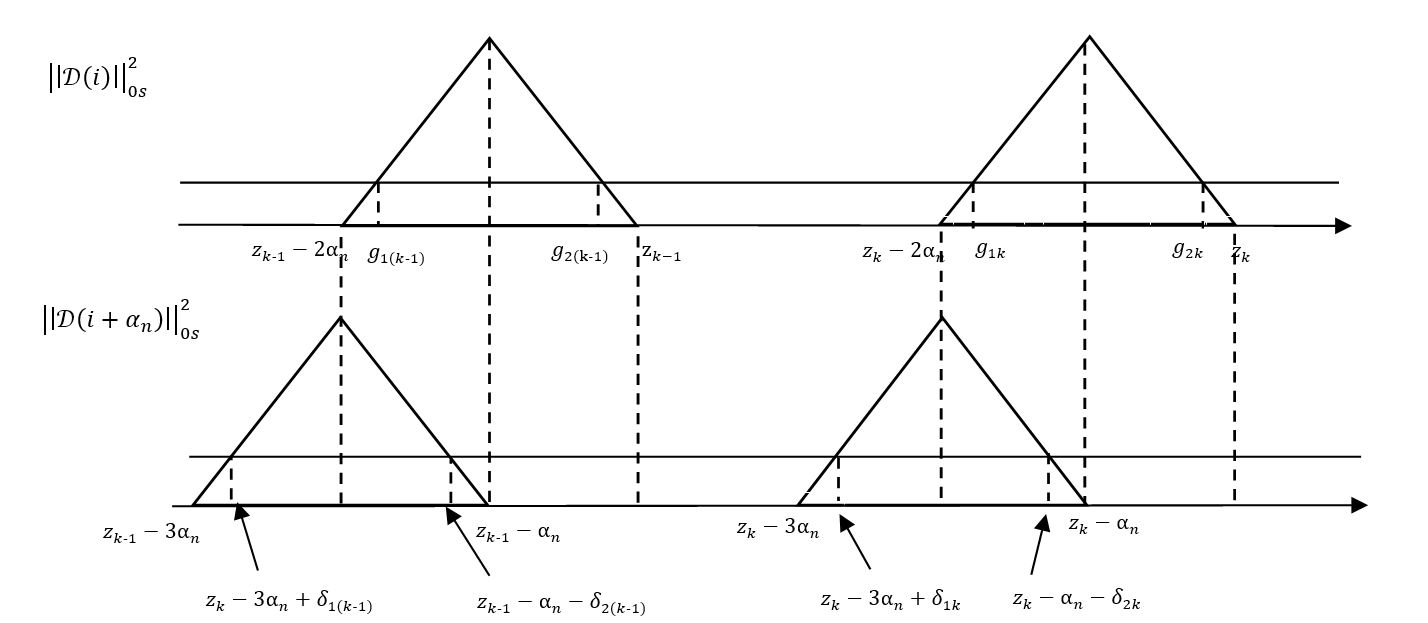}
   \includegraphics[width=0.7\textwidth]{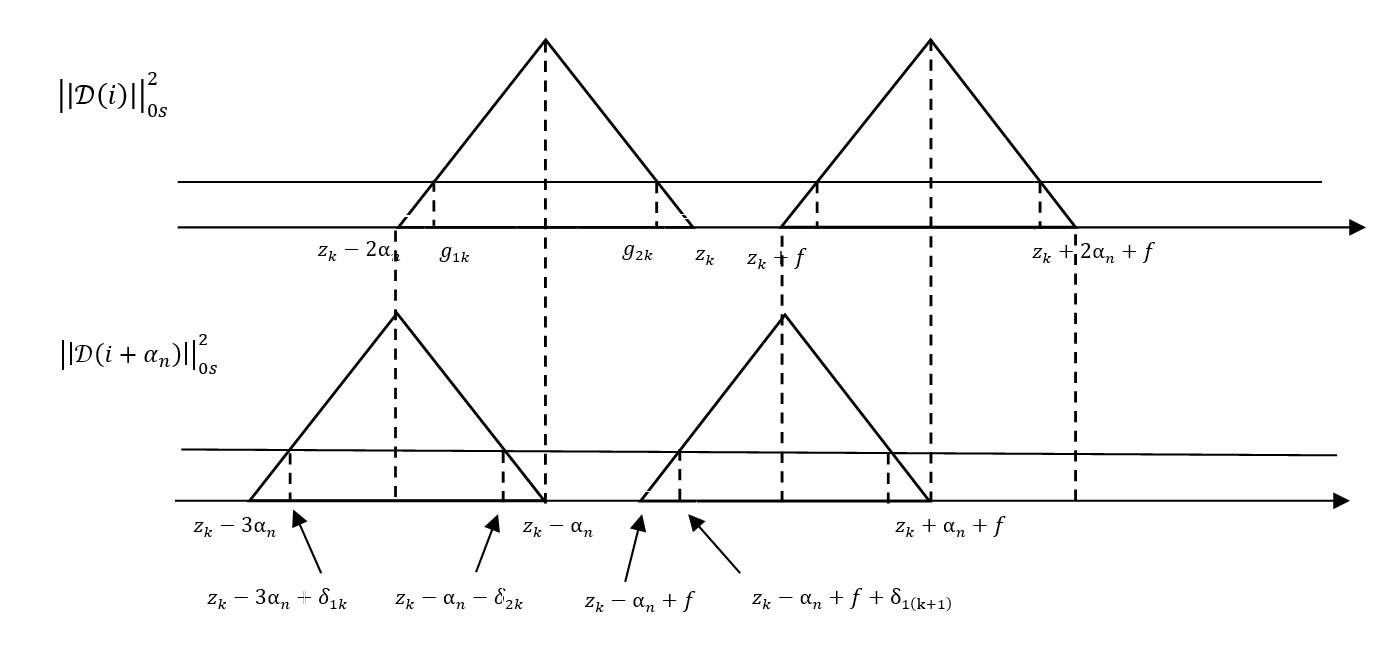}
\end{center}
  \caption{The upper plot is for the case where the distance between two true change points is not less than $3\alpha_n$, and the lower plot is about the case where the distance between two true change points is equal to $2\alpha_n+f,f\in[0,\alpha_n)$.}\label{tu1}
\end{figure}

Interestingly, as showed in Supplementary Materials, $||\mathcal{D}_n(i)||_s^2$, and $c_n(i)$ cannot respectively  converge to $||\mathcal{D}(i)||_{0s}^2$ and $c_n^{\star}(i)$ uniformly over all $i$, which further implies that the sample version $T_n(i)$  cannot  converge to  $T(i)$ uniformly over all $i$. Theorem~\ref{AoCRR} in Section~3 gives the details. Because of the different curve pattern of $T_n(\cdot)$ from that of $T(\cdot)$, we will suggest an algorithm to choose the intervals $(m_k(\tau), M_k(\tau))$ each containing only one local minimizer converging to those of $T(\cdot)$ in a certain sense.
The inconsistency occurs in the intervals where $||\mathcal{D}(i)||_{0s}^2$ and $||\mathcal{D}(i+\alpha_n)||_{0s}^2$ take small positive values. To help understand this intuitively, we give a toy example in Figure~\ref{tu1}. Theorem~\ref{AoCRR} in Section~3 further elaborates when the consistency of $T_n(\cdot)$ holds and when it does not, as well as how $T_n(\cdot)$ behaves when the consistency does not hold.\\

{\it  Choosing the intervals.} \, \, Motivated by Theorem~\ref{AoCRR} in Section~3,
%
we can, at the sample level, 
determine the disjoint intervals $( m_k(\tau),  M_k(\tau))$ for $k=1, \cdots, \hat K$ that are the estimated intervals of $(\tilde m_k(\tau), \tilde M_k(\tau))$'s as follows. First,  for a pre-determined threshold $0<\tau <1$, define $M_k(\tau)$ by
\begin{equation}\label{cri}
   T_n(M_k(\tau))<\tau, \quad T_n(M_k(\tau)+1)\geq \tau
\end{equation}
and let $m_k(\tau)=M_k(\tau)-\frac{2\sqrt{\tau}}{\sqrt{\tau}+1}\alpha_n$.
In this paper, we recommend $\tau=0.8.$  As stated in Theorem~\ref{AoCRR}, there might be some $M_{l}(\tau)$'s that correspond to false change points due to the inconsistency of $T_n(\cdot)$ in some sets in which we cannot determine the behavior of $T_n(\cdot)$. We call them the uncertain sets. Then we rule out those spurious $M_{l}(\tau)$'s when two conditions satisfy: $M_{l+1}(\tau)-M_{l}(\tau)\le 3\alpha_n/2$, and $T_n(M_{l}(\tau)-\alpha_n/2)\ge 1$, where $l\in \{l:T_n(M_l)\le \tau \text{ and } T_n(M_l+1)>\tau\}$.
 This is because  Theorem~\ref{AoCRR} offers two facts: 1). any $M_{l}(\tau)$ corresponding to a false change point has the distance to its nearest $M_{l+1}(\tau)$ corresponding to a true change point should be smaller than $(3-c)\alpha_n/2$, and the distance to $M_{l-1}(\tau)$ should be longer than $(3+c)\alpha_n/2$ for a small $c>0$; 2). $T_n(M_{l}(\tau)-\alpha_n/2) \to \infty$. \\

{\it  Identifying the locations of change points.} \, \, We can then search for local minimizers in the intervals   $(m_k(\tau), M_k(\tau))$'s with the remaining $M_k(\tau)$'s, and  use the minimizers plus $2\alpha_n-1$ as the estimators $\hat z_k$'s  of $z_k$'s. From Figure~\ref{population level}, we can see when the length $f$ is not big (shorter than $\alpha_n/2$ in theory), there might exist two or more local minima in one interval (at the population level, there would be exactly two local minima). The distance between one of them plus $2\alpha_n-1$  and $z_k$ might not be $o(\alpha_n)$. In other words, if this local minimizer is selected, the estimation consistency is destroyed. To avoid this problem, we
set,  for $1\leq k\leq \hat{K}$, $r_k := \max\{\arg\min_{i\in (m_k(\tau),M_k(\tau))}T_n(i)\}$ and the estimated location is defined as $\hat z_k=r_k+2\alpha_n-1$. 
\subsection{The mode-based SF distance and a signal statistic}

 From the above discussion, we can  define the mode-based signal screening distance and the corresponding signal statistic. Without loss of generality, we fix the last mode of the tensor to define the slice-wise MOSUM and the corresponding distance.
Define, for $  1\leq l\leq p_{\kappa}$,
\begin{align*}
  \mathcal{D}_l(i)=&\frac{1}{\alpha_n}\left(\sum_{j=i}^{i+\alpha_n-1}E(\mathcal{X}_{j,l})-\sum_{j=i+\alpha_n}^{i+2\alpha_n-1}E(\mathcal{X}_{j,l})\right),
\end{align*}
where for  $1\leq l\leq p_{\kappa}$, $\mathcal{X}_{j,l}\in R^{p_1\times p_2\times \cdots \times p_{\kappa-1}}$ are the $l$-th slices of the tensor after fixing the $\kappa$-th mode. The corresponding sample versions of $\mathcal{D}_{l}(i)$ and $T_{l}(i)$ are
$$\mathcal{D}_{nl}(i)=\frac{1}{\alpha_n}\left(\sum_{j=i}^{i+\alpha_n-1}\mathcal{X}_{j,l}-\sum_{j=i+\alpha_n}^{i+2\alpha_n-1}\mathcal{X}_{j,l}\right) \text{ and } T_{nl}(i)=\frac{||\mathcal{D}_{nl}(i)||_s^2+c_{nl}(i)}{||\mathcal{D}_{nl}(i+\alpha_n)||_s^2+c_{nl}(i)}.$$
Define $T^v(i)=\min_{1\leq l\leq p_{\kappa}}T_{l}(i).$  This sequence can identify structural changes at the population level.
 The final  signal statistic based on mode-based SF-distance is defined as $T_n^v(i)=\min_{1\leq l\leq p_{\kappa}}T_{nl}(i)$. 
 For each $l$, the definition of 
 $c_{nl}(i)$ is similar to $c_{n}(i)$ in 
 (\ref{cn}) with  $\mathcal S$ being replaced by, for each $1\le l\le p_{\kappa}$,  $\mathcal{S}_l=\left\{i:\sum_{t=1}^{\kappa-1}\sum_{i_t=1}^{p_t}I\left(\mathcal{D}_{nl,i_1i_2\cdots i_{\kappa}-1}^2(i)>l_n(s)\right)> 0 \right\}.$
 Theorem~\ref{th2} in Section~3 offers the estimation consistency of the locations and number of the change points. 
 \begin{figure}[h]
\begin{center}
   \includegraphics[width=0.7\textwidth]{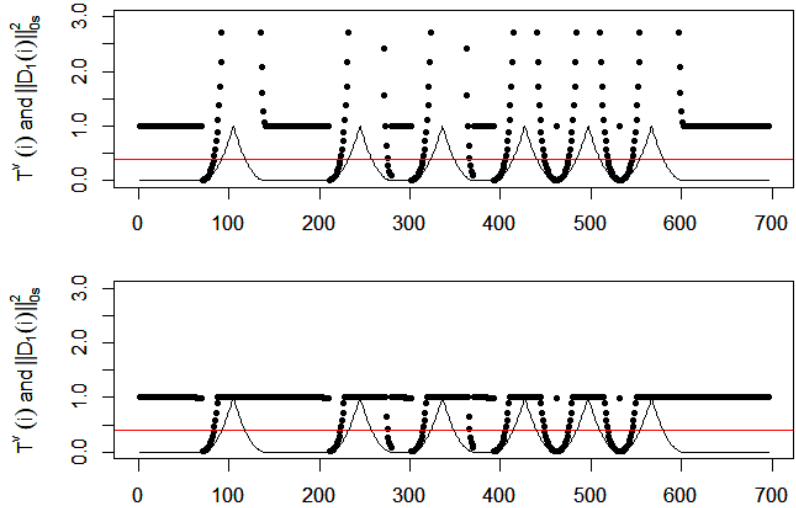}
\end{center}
  \caption{the curves of $||\mathcal{D}_1(\cdot)||_{0s}^2$ and $T^v(\cdot)$. The dashed line is for $||\mathcal{D}_1(\cdot)||_{0s}^2$ and the dot line for $T^v(\cdot)$. The upper plot presents the curve when all elements of the vector have changes at the same locations; the lower plot is for the curves when there is at least an element of the vector having changes at different locations.}\label{populationvector}
\end{figure}
Write $T^v(\cdot)$ and $||\mathcal{D}_l(\cdot)||_{0s}^2$ as the corresponding functions at the population level.   From the two plots of Figure~\ref{populationvector} compared with Figure~\ref{population level}, we can see that  {the curve of $T^v(i)$  is different from that of $T(\cdot)$.  The top plot shows that if the change points occur in all slices, the curve of $T^v(\cdot)$ has the same patten as  that of $T(\cdot)$; otherwise, they are different as shown in the bottom plot.}

Similarly, to those with $T_n(\cdot)$, we also choose the relevant intervals to identify the locations of the local minima. 

{\it  Choosing the intervals.} \, \, By the results of Theorem~\ref{AoCRR-modebased} in Section~3,
we  
similarly determine the disjoint intervals $(m_k^v(\tau),  M_k^v(\tau))$ for $k=1, \cdots, \hat K$ as follows. First, for a pre-determined threshold $0<\tau <1$, define $M_k^v(\tau)$ by
\begin{equation}\label{criv}
   T_n^v(M_k^v(\tau))<\tau, \quad T_n^v(M_k^v(\tau)+1)\geq \tau
\end{equation}
and let $m_k^v(\tau)=M_k^v(\tau)-\frac{2\sqrt{\tau}}{\sqrt{\tau}+1}\alpha_n$.
As $T_n^v(\cdot)$ is the minimum of $T_{nl}(\cdot)$, the values tend to be small at the sample level. We then, in this paper, recommend $\tau=0.4.$  As stated in Theorem~\ref{AoCRR-modebased} in the next section,  the inconsistency of $T_n^v(\cdot)$ occurs in the uncertain sets in  which  we cannot determine what behavior of $T_n^v(\cdot)$ is. Then, there might also exist some $M_{g}$'s that correspond to false change points. We rule out those $M_g$'s when  $M_{g+1}(\tau)-M_{g}(\tau)\le 3\alpha_n/2$, where $g\in \{g:T_n^v(M_g)<\tau \text{ and } T_n^v(M_g+1)\geq \tau\}$. This is a different algorithm from that with $T_n(\cdot)$. It is because Theorem~\ref{AoCRR-modebased} offers the fact:  any $M_g(\tau)$ corresponding to a possible false change point has the distance to its nearest $M_{g+1}(\tau)$ corresponding to a true change point should be smaller than $(3-c)\alpha_n/2$, and the distance to $M_{g-1}(\tau)$ should be longer than $(3+c)\alpha_n/2$ for a small $c>0$. But we no longer have the property that $T_n^v(M_{g}(\tau)-\alpha_n/2) \to \infty$. See it intuitively  from Figure~\ref{populationvector}.\\

{\it  Identifying the locations of change points.} \,  This step can be similar to that with $T_n(\cdot)$ by  identifying the local minimizer $r_k :=\max\{\arg\min_{i\in (m_k^v(\tau),M_k^v(\tau))}T_n^v(i)\}$ for $1\leq k\leq \hat{K}$ and the estimated location is defined as $\hat z_k=r_k+2\alpha_n-1$.

\section{The asymptotic properties}
This section gives the asymptotic properties of the proposed signal statistics, the estimated locations $\hat z_k$, and the estimated number $\hat K$ of change points. As $T_n^v(\cdot)$ has different asymptotic behaviors from $T_n(\cdot)$, we then also give the results about $T_n(\cdot)$ first.

\subsection{The asymptotic properties of the SF-distance-based  statistic}
Introduce some notations to be used in the following theorems. Denote ${\rm{Var}}({\rm{vec}}(\mathcal{E}))=\colon \boldsymbol{\Sigma}=\left(\sigma_{j_1j_2}\right), 1\leq j_1,j_2\leq p$, $p=\prod_{l=1}^{\kappa}p_l$ and $p_{-\kappa}=\prod_{l=1}^{\kappa-1}p_l$, where ${\rm{vec}}(\cdot)$ represents the vectorization of a tensor. Let $\boldsymbol{\eta}=\boldsymbol{\Sigma^{-\frac{1}{2}}}{\rm{vec}}(\mathcal{E})$, $\boldsymbol{\eta}^{(l)}=\boldsymbol{\Sigma^{-\frac{1}{2}}}{\rm{vec}}(\mathcal{E}_l)$, 
 $a_n=\frac{(logn)^{1+2\epsilon}}{(\log n)^{\rho_1}\alpha_n\left(\log np\right)}$, and $b_n=\frac{(a_n\alpha_n-2C^{\prime})^2}{(\log n)^{\rho_2}\log[(K+1)pn](K+1)^2}$ for three positive constants $\epsilon$, $\rho_1$ and $\rho_2$.

\begin{proposition}\label{con1}
 Assume that $\mathcal{X}_i-E(\mathcal{X}_i)$ are independent random tensors.
 \begin{itemize}
   \item [(i)] Under  Condition~1 in Supplementary Materials, when $\frac{n^2p}{\alpha_n^4b_n^2}\to 0$,
 we have the following result for the SF-distance-based MOSUM:
\begin{align}\label{l1}
   \max_{i}\max_{i_1,i_2,\cdots,i_{\kappa}}\left\vert \mathcal{D}_{n,i_1i_2\cdots i_{\kappa}}(i)-\mathcal{D}_{i_1i_2\cdots i_{\kappa}}(i)\right\vert =O_p\left(\frac{(\log n)^{\frac{1}{2}+\epsilon}}{\sqrt{\alpha_n}}\right).
\end{align}
\item [(ii)]  Assume that the last mode's indexes of $\mathcal{X}_i$ are independent  and the elements $\eta_{ij}$ of $\boldsymbol \eta_i$ for $i=1, 2, \cdots, n;j=1, 2, \cdots, p$ are independent and identically distributed. Suppose that Conditions~2 and~3 in Supplementary Materials hold and $\frac{\alpha_n^5\log\left(np\right)}{n^{4}}=o(1)$.
If $\left[E|\eta_{11}|^q\right]^{1/q}<\infty \text{ and } \frac{(\log(np))^{3q/2}np}{\alpha_n^{q/2}}=o(1)$ hold for some $q\geq 4$, or $E(\exp^{a_0\eta_{11}})<\infty \text{ and }\frac{(\log(np))^{\max\{7,2(1+\beta)/\beta\}}}{\alpha_n}=o(1)$  for some $a_0>0$ hold, then we have the following result for the SF-distance-based MOSUM:
\begin{align}\label{l2}
   \max_{i}\max_{i_1,i_2,\cdots,i_{\kappa}}\left\vert \mathcal{D}_{n,i_1i_2\cdots i_{\kappa}}(i)-\mathcal{D}_{i_1i_2\cdots i_{\kappa}}(i)\right\vert =O_p\left(\frac{\left(\log\left(np\right)\right)^{1/2+\epsilon}}{\sqrt{\alpha_n}}\right).
   \end{align}
 \end{itemize}
 \end{proposition}
For notational simplicity, denote $\epsilon_n=\frac{(\log n)^{\frac{1}{2}+\epsilon}}{\sqrt{\alpha_n}}$ in Part (i) and $\epsilon_n=\frac{\left(\log\left(np\right)\right)^{1/2+\epsilon}}{\sqrt{\alpha_n}}$ in Part (ii).
\begin{rema}
(Allowed dimension)
 This proposition shows the critical differences between Parts (i) and (ii).  Particularly,
under additional conditions on $\eta_{ij}$, such as the IID condition as well as other conditions on its moment and tail probability (see, e.g., Theorem 1 in \cite{chen2021inference}), we can see from Part (ii) that the divergence rate of $p$ can be fast.
  While when these conditions are relaxed as in Part (i), we still have the estimation consistency of $\mathcal{D}(i)$ with a slower divergence rate of $p$.

In Part (i),  the formulas of $a_n$ and $b_n$ give that
 $$\frac{n^2p}{\alpha_n^4b_n^2}=\frac{n^2p(\log n)^{4\rho_1+2\rho_2}\left[\log \left(np\right)\right]^4\left\{\log [(K+1)np]\right\}^2(K+1)^4}{\alpha_n^4\left[(\log n)^{1+2\epsilon}-2C^{\prime}(\log n)^{\rho_1}\log \left(np\right)\right]^4}\to 0.$$
Take $\epsilon=\frac{\rho_1}{2}+\frac{\rho_2}{4}+1$, 
 then $\frac{n^2p\left\{\log[(K+1)n]\right\}^2\left[\log (np)\right]^{4}(K+1)^4}{\alpha_n^4[\log n]^6}\to 0$. When the number of change points is fixed and $\alpha_n=n^{\frac{1}{2}+\nu}$ with $\nu\in (0,\frac{1}{2})$, we can see that $p$ can be  close to the order $n^{4\nu}$ and  can be close to  $n^2$ at most. More importantly, we do not require that $\mathcal{X}_i-E(\mathcal{X}_i)$ should be identically distributed or specify that the data come from some specific distribution.
 %

In Part (ii), under the finite moment conditions, $p$ can be at a polynomial order of $n$, depending on $q$. The larger $q$ allows the larger dimension $p$. Under the exponential tail condition, $p$ can be exponential in $n$.
\end{rema}

Define $\mathcal{G}=\left\{i: \exists i_1,i_2,\cdots,i_{\kappa}, \mathcal{D}_{i_1i_2\cdots i_{\kappa}}^2(i)>0 \text{ and } \mathcal{D}_{n,i_1i_2\cdots i_{\kappa}}^2(i)\leq  l_n(s)\right\},$
and $(\mathcal{G}+\alpha_n)= \{i: i-\alpha_n \in \mathcal{G} \}$. Further, let $g_{1k}=\arg\min_{j\in \mathcal{G},j+1\notin \mathcal{G}}|j-(z_k-2\alpha_n)|$ and $g_{2k}=\arg\min_{j\in \mathcal{G},j-1\notin \mathcal{G}}|j-z_k|$ for $k=1,2,\cdots, K$.
\begin{proposition}\label{com-con}
   Under the conditions in Proposition~\ref{con1}, we have
\begin{itemize}
  \item [(i)] $\max_{i\in \mathcal{G}^c}\big\vert ||\mathcal{D}_{n}(i)||_s^2-||\mathcal{D}(i)||_{0s}^2\big\vert=O_p(\epsilon_n)$.
\item [(ii)] Let $\delta_{1k}=g_{1k}-(z_k-2\alpha_n)$ and $\delta_{2k}=z_k-g_{2k}$. If Condition~6 holds, then $\max_{1\leq k\leq K}\delta_{1k}=o_p(\alpha_n)$ and $\max_{1\leq k\leq K}\delta_{2k}=o_p(\alpha_n)$.
\item [(iii)] Let $\mathcal{M}_n=\{i:T_n(i)\leq M_n(i)\}$, where $M_n=o\left(\frac{c_n(i)}{\epsilon_n}\right)$. If Condition~2 holds, $\max_{i\in \mathcal{G}^c\cap (\mathcal{G}+\alpha_n)^c\cap \mathcal{M}_n}\left\vert T_n(i)-T(i)\right\vert=o_p(1)$.
 \end{itemize}
 \end{proposition}
  The above results are the basis for obtaining the consistency of the estimated locations and number of change points when the signal statistic $T_n(\cdot)$ is used. There may exist some intervals with length $o(\alpha_n)$ such that $||\mathcal{D}_{n}(i)||_s^2$ does not converge to $||\mathcal{D}(i)||_{0s}^2$, and similarly for $T_n(\cdot)$.  

  We will, in the following theorem, give some detailed analysis of the inconsistency and some more asymptotic behaviors of $T_n(\cdot)$ in the set $\mathcal{G}\cup (\mathcal{G}+\alpha_n).$ The proof is included in Supplementary Materials. Importantly, we will show that this set is sufficiently small and does not influence the choice of $M_k(\tau)$ and the consistency of $\hat{z}_k$.

 Note that the inconsistencies of $\mathcal{D}_n(\cdot)$, $\mathcal{D}_n(\cdot+\alpha_n)$ and $c_n(\cdot)$ in the set $ \mathcal{G}\cup (\mathcal{G}+\alpha_n)$ could cause the inconsistency of $T_n(\cdot)$.
  From the analysis to the set $\mathcal{G}^c$ in the Part (i) of Proposition~\ref{com-con}, we can know that $\mathcal{G}^c\subset \{i:c_n(i)=c_n^{\star}(i)\}$ with a probability going to one. In other words,  $\{i:c_n(i)\not =c_n^{\star}(i)\}\subset \mathcal{G}$ with a probability going to one.

Divide $\mathcal{G}$ into two sets $\mathcal{L}=\mathcal{S}^{\star}\cap \mathcal{S}^c$ and $\mathcal{G}\backslash \mathcal{L}$ where  $\mathcal{L}$ can be written as
\begin{equation}\label{L-d}
  \left\{i: \exists i_1^0,i_2^0,\cdots,i_{\kappa}^0,\mathcal{D}_{i_1^0i_2^0\cdots i_{\kappa}^0}^2(i)>0 \text{ and } \forall i_1,i_2,\cdots,i_{\kappa}, \mathcal{D}_{n,i_1i_2\cdots i_{\kappa}}^2(i)\leq l_n(s) \right\}.
\end{equation}
Note that  $c_n(i)=s_1n(\log (n))^{\nu}\epsilon_n \to \infty, $ for $ i\in \mathcal{L}$  defined in (\ref{L-d}), and $c_n(i)=\frac{s_1(\log(n))^{\nu}\epsilon_n}{I(i\in \mathcal{S})+\frac{1}{n}}\to 0, $ for $i\in \mathcal{G}\backslash \mathcal{L}$.
Further, the set $\mathcal{L}$ also consists of $2K$ disjoint intervals, each containing in the corresponding interval in $\mathcal{G}$. Write $l_{1k}=\arg\min_{j\in \mathcal{L},j+1\notin \mathcal{L}}|j-(z_k-2\alpha_n)|$ and $l_{2k}=\arg\min_{j\in \mathcal{L},j-1\notin \mathcal{L}}|j-z_k|$ for $k=1,2,\cdots, K$. Define $l_{1k}-(z_k-2\alpha_n)=\colon \Delta_{1k}$ and $z_k-l_{2k}=\colon \Delta_{2k}$. Since $\mathcal{L}\subset \mathcal{G}$, we have $\Delta_{1k}\leq \delta_{1k}$ and $\Delta_{2k}\leq \delta_{2k}$.
\begin{theorem}
\label{AoCRR}  
Under the conditions in Proposition~\ref{com-con}, Conditions~7 and~8 in Supplementary Materials, we have the following results hold with a probability going to $1$.

Case 1. When the distance between two adjacent change points is not less than $3\alpha_n$,
\begin{itemize}
  \item [1.] $\min_{k}\min_{i\in (z_k-3\alpha_n, z_k-3\alpha_n+\delta_{1k}]}T_n(i)\to 1$;
  \item [2.] $\min_{k}\min_{i\in(z_k-2\alpha_n,l_{1k}]}T_n(i)\to 1$ and $\max_{k}\max_{i\in(l_{1k},g_{1k}]}T_n(i)\to 0$;
  \item [3.]  $\lim \inf_n\min_{k}\min_{i\in[z_k-\alpha_n-\delta_{2k},z_k-\alpha_n)}T_n(i)\ge 1$;
  \item [4.] $\liminf_n\min_{k}\min_{i\in[g_{2k},z_k)}T_n(i)\ge 1.$
\end{itemize}

Case 2.When the distance between two adjacent change points is equal to $2\alpha_n+f,f\in [0,\alpha_n)$,
 \begin{itemize}
  \item  [1.]  $\min_k\min_{i\in (z_k-2\alpha_n,l_{1k}]}T_n(i)\to 1$ and $\max_k\max_{i\in(l_{1k},g_{1k}]}T_n(i)\to 0;$
  \item  [2.] $\lim \inf_n \min_k\min_{i\in [z_k-\alpha_n-\delta_{2k},z_k-\alpha_n)}T_n(i)\to1;$
  \item  [3.] $\lim\inf_n\min_k\min_{i\in(z_k-\alpha_n+f, z_k-\alpha_n+f+\delta_{1(k+1)}]\backslash [g_{2k},z_k)}T_n(i)\ge 1;$
  \item  [4.] Recall the definition of $M_k(\tau)$ in Subsection~2.2. Define $c_n=\frac{(\log n)^{\nu}\epsilon_n}{1+\frac{1}{n}}$.
  \begin{itemize}
    \item [(i)] For $f\in \left[0,\alpha_n-\delta_{2k}-\frac{\alpha_n \sqrt{A_2}}{\min_k\min_{\mathcal{H}_k}\gamma_{i_1i_2\cdots i_{\kappa}}^k}\right]$,  $\max_k\max_{i\in [g_{2k},l_{2k})}T_n(i)\to 0,$
        where $A_2=\left[\left(1-\frac{\sqrt{\tilde{l}_n(s)}}{\min_k\min_{\mathcal{H}_k}\gamma_{i_1i_2\cdots i_{\kappa}}^k}\right)\min_k\min_{\mathcal{H}_k}\gamma_{i_1i_2\cdots i_{\kappa}}^k\right]^2.$
    \item [(ii)] For $f\in \left[\alpha_n-\Delta_{2k}-\frac{\alpha_n \sqrt{c_n}}{(\log n)^{\frac{1}{8}}\min_k\min_{\mathcal{H}_k}\gamma_{i_1i_2\cdots i_{\kappa}}^k},\alpha_n\right)$,
      $\lim\inf_n\min_k\min_{i\in [g_{2k},l_{2k})}T_n(i)\geq 1; $
    \item [(iii)] For $f\in \left(\alpha_n-\delta_{2k}-\frac{\alpha_n \sqrt{A_2}}{\min_k\min_{\mathcal{H}_k}\gamma_{i_1i_2\cdots i_{\kappa}}^k},\alpha_n-\Delta_{2k}-\frac{\alpha_n \sqrt{c_n}}{(\log n)^{\frac{1}{8}}\min_k\min_{\mathcal{H}_k}\gamma_{i_1i_2\cdots i_{\kappa}}^k}\right)$, the set $[g_{2k},l_{2k})$ is an uncertain set in the sense that  the convergence is unclear. But,
$\lim\inf_n\min_k\min_{i\in[g_{2k},l_{2k})}T_n(i-\frac{\alpha_n}{2})\to \infty$ and \\ $\lim\sup_n\max_k\max_{i\in [z_k+f,z_k+f+\frac{\alpha_n}{2}]}T_n(i-\frac{\alpha_n}{2})\leq 1.$ Further, there is a small value $c>0$ such that
$$\max_k\max_{i\in [g_{2k},l_{2k})}\left\vert \tilde{M}_{k+1}(\tau)-i\right\vert < \frac{(3-c)\alpha_n}{2} \text{ and }\min_k\min_{i\in [g_{2k},l_{2k})}\left\vert i-\tilde{M}_k(\tau)\right\vert> \frac{(3+c)\alpha_n}{2}.$$
   \item [(iv)] $\min_k\min_{i\in[l_{2k},z_k)}T_n(i)\to 1.$
   \end{itemize}
\end{itemize}
  \end{theorem}
  \begin{rema}
 The analysis of the results in this theorem is rather tedious. To help understand the results intuitively, we give a toy example to present  the set $\mathcal{G}$ in which $\|\mathcal{D}_n(i)\|_s^2$ may not converge to $\|\mathcal{D}(i)\|_{0s}^2$, and the set $(\mathcal{G}+\alpha_n)$. The plots in Figure~\ref{tu1}  respectively
 show the curves in the cases where the distance between two adjacent true change points is not less than $3\alpha_n$ and equal to $2\alpha_n+f,f\in [0,\alpha_n)$. We can obviously see that for each $k$, there are four disjoint intervals in $\mathcal{G}$  and $(\mathcal{G}+\alpha_n)$  discussed in the theorem such that $T_n(\cdot)$ may not converge to $T(\cdot)$. When the distance between two adjacent change points is shorter than $3\alpha_n$, Part 4 in Case 2 shows that the behavior of $T_n(\cdot)$ in some small intervals is very complicated.  But the above analysis will help choose the intervals, each containing only one change point in Propostion 6 in Supplementary Materials.
\end{rema}

Define $\hat z_k=0$ if $\hat K <k \le K$.

\begin{theorem}\label{th1}
     Under the conditions in Proposition~\ref{com-con},
     when Conditions~6 and 7 hold in Supplementary Materials, $\alpha_n^{\star}\geq 2\alpha_n$ and $\frac{n}{K\alpha_n^{\star}}\to \infty$, we have $Pr(\hat{K}=K)\to 1,$
    and the estimators $\{\hat z_1, \hat z_2,\cdots, \hat z_{\hat K}\}$ have $\Pr\left\{\max_{1\leq k\leq  K}\left\vert\frac{\hat{z}_k-z_k}{\alpha_n}\right\vert<\epsilon\right\}\to 1$ for every $\epsilon>0.$
    \end{theorem}
%

\begin{rema}

 Theorem \ref{th1} shows the estimation accuracy of the change point location is of order $o_p(\alpha_n)$ with $\alpha_n\gg n^{\frac{1}{2}}p^{\frac{1}{4}}$ in the non-i.i.d. error case (see Proposition \ref{con1} (i)) or $o_p(\alpha_n)$ with $\frac{\alpha_n^5\log (np)}{n^4}\to 0$ in the i.i.d. error case (see, Proposition~\ref{con1} (ii)). In Supplementary Material, Condition~7 states the minimum signal strength in this paper.

 Some comparison with existing results in the literature is in order. \cite{wang2021optimal} focused on dynamic networks with binary elements, 
 which can be regarded as an order-two tensor. In the i.i.d. error cases its estimation accuracy can achieve the order $o(\alpha_n^{\star})$ that is the minimum length between two change points. The constraint $\alpha_n^{\star}\ge 2 \alpha_n$ we require is not assumed in their results, but some other conditions are required.
For example,  as they commented in their paper (see, \cite{wang2021optimal}, page 210), the denseness in their result requires that 
more than $O(\sqrt{p})$ elements must change in the same location of any change point.  Our method does not need this condition. Further, how to estimate the sparsity is still an unsolved problem in their method.
  \cite{wang2018high}  can achieve the accuracy order  $o(\sqrt{n})$ that is better than our accuracy order of $o(\alpha_n)$  in the current paper; but mainly  under sparse models and  normality.

\end{rema}

\subsection{The asymptotic properties of the MSF-distance-based  statistic}
We now  analyse $T_n^v(\cdot)$. Define $$\mathcal{G}_l=\left\{i: \exists i_1,i_2,\cdots,i_{\kappa-1}, \mathcal{D}_{l,i_1i_2\cdots i_{\kappa-1}}^2(i)>0 \text{ and } \mathcal{D}_{nl,i_1i_2\cdots i_{\kappa-1}}^2(i)\leq  l_n(s)\right\}.$$
Proposition~\ref{com-con} also shows the uniform consistency of  $\|\mathcal{D}_{nl}(i)\|_s^2$ to $\|\mathcal{D}_l(i)\|_{0s}^2$ over the set $\mathcal{G}_l^c$, while may not be  for  $i\in \mathcal{G}_l$. Thus, according to Proposition~\ref{AoCRR}, $T_{nl}(i)$ maintains the uniform consistency over $\mathcal{G}_l^c\cap (\mathcal{G}_l+\alpha_n)^c\cap \mathcal{M}_{nl}$, where $(\mathcal{G}_l+\alpha_n)= \{i: i-\alpha_n \in \mathcal{G}_l\}$, $\mathcal{M}_{nl}=\{i:T_n(i)\leq M_{nl}(i)\}$ and $M_{nl}(i)=o\left(\frac{c_{nl}(i)}{\epsilon_n}\right)$.

Recall the definitions of $\mathcal{S}_l$ in Subsection~2.4. For notational convenience, write $\mathcal{L}_l$ as $\mathcal{S}_l^c\cap \{i:||\mathcal{D}_l(i)||_{0s}^2>0\}=\colon\mathcal{S}_l^c\cap \mathcal{S}_l^{\star}$.  Similar to the analysis for the signal statistic SFD before Theorem~\ref{AoCRR}, the existence of $\mathcal{L}_l$  also influences the uniform consistency of $T_{nl}(i)$. As $\mathcal{L}_l\subseteq\mathcal{G}_l$ that is proved previously,  we now use some similar arguments that have been used for each $T_{n}(i)$ to obtain the uniform consistency of $T_n^v(\cdot)$ over $\mathcal{G}^{vc}\cap(\mathcal{G}^v+\alpha_n)^{c}\cap \mathcal{M}_n^v$, where $\mathcal{G}^v=\cup_{l=1}^{p_{\kappa}} \mathcal{G}_l$, $(\mathcal{G}^v+\alpha_n)= \{i: i-\alpha_n \in \mathcal{G}^v\}$ and  $\mathcal{M}_n^v=\cap_{l=1}^{p_{\kappa}} \mathcal{M}_{nl}$.

Parallel to Theorem~\ref{AoCRR},  Theorem~\ref{AoCRR-modebased} below shows that $T_n^v(\cdot)$ has the uniform consistency over $\mathcal{G}^{vc}\cap(\mathcal{G}^v+\alpha_n)^{c}\cap \mathcal{M}_n^v$, and is inconsistent when $i\in \mathcal{G}^{v}\cup (\mathcal{G}^{v}+\alpha_n)\cap \mathcal{M}_n^{vc}$, but  keeps  similar properties of $T_n(i),i\in \mathcal{G}\cup (\mathcal{G}+\alpha_n)$ except for Part 4(iii) in Case 2 of Theorem~\ref{AoCRR}. Again, the inconsistency of $T_n^v(i)$  does not influence choosing the disjoint intervals, each containing only one change point, as stated in Proposition~8 in Supplementary Materials.

Now we give some notations that will be used in the following results.
Similar to the definitions of $g_{1k}$ and $g_{2k}$ before Proposition~\ref{com-con}, we define $g_{l,1k}=\arg\min_{j\in \mathcal{G}_l,j+1\notin \mathcal{G}_l}|j-(z_k-2\alpha_n)|$ and $g_{l,2k}=\arg\min_{j\in \mathcal{G}_l,j-1\notin \mathcal{G}_l}|j-z_k|$ for $k=1,2,\cdots, K$. Then $\max_{1\leq l\leq p_{\kappa}}g_{l,1k}=\colon g_{1k}^v$ and $\min_{1\leq l\leq p_{\kappa}}g_{l,2k}=\colon g_{2k}^v$ are the right- and left-hand ending point of $\mathcal{G}^v$ respectively. The lengths of $2K$ disjoint intervals in $\mathcal{G}^v$ are defined as $g_{1k}^v-(z_k-2\alpha_n)=\colon \delta_{1k}^v$ and $z_k-g_{2k}^v=\colon \delta_{2k}^v, k=1,2,\cdots,K$.

From the proof of Part (ii) in Proposition~\ref{com-con} presented in Supplementary Materials, we know that for any $l$, $g_{l,1k}-(z_k-2\alpha_n)=o(\alpha_n)$ and  $z_k-g_{l,2k}=o(\alpha_n)$. Notice that the bound $\sqrt{\tilde{l}_n(s)}\alpha_n/\min\min_{\mathcal{H}_k}\gamma_{i_1i_2\cdots i_{\kappa}}^k$ is free from $l$. Thus, we have $\delta_{1k}^v=o(\alpha_n)$ and $\delta_{2k}^v=o(\alpha_n)$. For $k=1,2,\cdots, K$, write $l_{1k}^v=\min_{1\leq l\leq p_{\kappa}}\left\{\arg\min_{j\in \mathcal{L}_l,j+1\notin \mathcal{L}_l}|j-(z_k-2\alpha_n)|\right\}\text{ and } l_{2k}^v=\max_{1\leq l\leq p_{\kappa}}\left\{\arg\min_{j\in \mathcal{L}_l,j-1\notin \mathcal{L}_l}|j-z_k|\right\}.$ Define $l_{1k}^v-(z_k-2\alpha_n)=\colon \Delta_{1k}^v$ and $z_k-l_{2k}^v=\colon \Delta_{2k}^v$. Due to $\mathcal{L}_l\subseteq \mathcal{G}_l$, we have $\Delta_{1k}^v\leq \delta_{1k}^v$ and $\Delta_{2k}^v\leq \delta_{2k}^v$. 

Similar to the definition of $M_k(\tau)$ and $m_k(\tau)$, let
        $M_k^v(\tau)$ and $m_k^v(\tau)$ be defined by $T_n(M_k^v(\tau))<\tau$, $T_n(M_k^v(\tau)+1)\geq \tau,\tau \in (0,1)$ and $m_k^v(\tau)=M_k^v(\tau)-\frac{2\sqrt{\tau}}{\sqrt{\tau}+1}\alpha_n$. $(m_{kl}(\tau),M_{kl}(\tau))$ and $(\tilde{m}_{kl}^v(\tau),\tilde{M}_{kl}^v(\tau))$ are defined by $T_n^v(M_{kl}^v(\tau))<\tau,T_n^v(M_{kl}^v(\tau)+1)\geq \tau,$ $T^v(\tilde{M}_{kl})<\tau, T^v(\tilde{M}_{kl}^v(\tau)+1)\geq \tau$, and $m_{kl}^v(\tau)=M_{kl}^v(\tau)-\frac{2\sqrt{\tau}}{\sqrt{\tau}+1}\alpha_n$.

  \begin{theorem}\label{AoCRR-modebased}
    Under the conditions in Proposition~\ref{com-con}, we have the following results with a probability going to one.

$\bullet$ Additionally, assume Condition~8 in Supplementary Material holds.

Case 1. When the distance between two adjacent change points is not less than $3\alpha_n$, for any $k=1,2,\cdots,K$, the union of all the four sets in following Parts 1-4 for all $k=1,2,\cdots,K$ are equal to $ \mathcal{G}^v\cup (\mathcal{G}+\alpha_n)^v$.
\begin{itemize}
  \item  [1.] $\min_k\min_{i\in (z_k-3\alpha_n, z_k-3\alpha_n+\delta_{1k}^v]}T_n^v(i)\to 1$.
  \item  [2.] $\min_k\min_{i\in(z_k-2\alpha_n,l_{1k}^v]}T_n^v(i)\to 1$ and $\max_k\max_{i\in(l_{1k}^v,g_{1k}^v]}T_n^v(i)\to 0;$
  \item  [3.] $\lim \inf_n \min_k\min_{i\in[z_k-\alpha_n-\delta_{2k}^v,z_k-\alpha_n)}T_n^v(i)\ge 1;$
  \item  [4.] $\lim\inf_n\min_k\min_{i\in [g_{2k}^v,z_k)}T_n^v(i)\ge 1.$
\end{itemize}

Case 2. When the distance between two adjacent change points is equal to $2\alpha_n+f,f\in [0,\alpha_n)$, the union of all the four sets in following Parts 1-4 for all $k=1,2,\cdots,K$ are equal to $ \mathcal{G}^v\cup (\mathcal{G}+\alpha_n)^v$.
 \begin{itemize}
  \item  [1.]  $\min_k\min_{i\in (z_k-2\alpha_n,l_{1k}^v]}T_n^v(i)\to 1$ and $\max_k\max_{i\in(l_{1k}^v,g_{1k}^v]}T_n^v(i)\to 0;$
  \item  [2.] $\lim\inf_n\min_{i\in [z_k-\alpha_n-\delta_{2k}^v,z_k-\alpha_n)}T_n^v(i)\ge 1;$
  \item  [3.] $\lim\inf_n\min_{i\in(z_k-\alpha_n+f, z_k-\alpha_n+f+\delta_{1(k+1)}^v)\backslash[g_{2k}^v,z_k)}T_n^v(i)\ge 1;$
  \item  [4.]  Recall the definition of $M_k^v(\tau)$ before this theorem.
  \begin{itemize}
    \item [(i)] For $f\in \left[0,\alpha_n-\delta_{2k}^v-\frac{\alpha_n \sqrt{A_2}}{\min_k\min_{\mathcal{H}_k}\gamma_{i_1i_2\cdots i_{\kappa}}^k}\right]$,
      $\max_k\max_{i\in [g_{2k}^v,l_{2k}^v)}T_n^v(i)\to 0;$
    \item [(ii)]   For $f\in \left[\alpha_n-\Delta_{2k}^v-\frac{\alpha_n \sqrt{c_n}}{(\log n)^{\frac{1}{8}}\min_k\min_{\mathcal{H}_k}\gamma_{i_1i_2\cdots i_{\kappa}}^k},\alpha_n\right)$,
      $\lim\inf_n\min_k\min_{i\in [g_{2k}^v,l_{2k}^v)}T_n^v(i)\geq 1; $
    \item [(iii)]  For $f\in \left(\alpha_n-\delta_{2k}^v-\frac{\alpha_n \sqrt{A_2}}{\min_k\min_{\mathcal{H}_k}\gamma_{i_1i_2\cdots i_{\kappa}}^k},\alpha_n-\Delta_{2k}^v-\frac{\alpha_n \sqrt{c_n}}{(\log n)^{\frac{1}{8}}\min_k\min_{\mathcal{H}_k}\gamma_{i_1i_2\cdots i_{\kappa}}^k}\right),$
      $$\max_k\max_{i\in [g_{2k},l_{2k})}\left\vert \tilde{M}_{k+1}^v(\tau)-i\right\vert < \frac{(3-c)\alpha_n}{2} \text{ and } \min_k\min_{i\in [g_{2k},l_{2k})}\left\vert i-\tilde{M}_k^v(\tau)\right\vert> \frac{(3+c)\alpha_n}{2}.$$ 
   \item [(iv)] $\min_k\min_{i\in[l_{2k}^v,z_k)}T_n^v(i)\to 1.$
 \end{itemize}
 \end{itemize}
$\bullet$ Furthermore, $\max_{i\in \mathcal{G}^{v c}\cap (\mathcal{G}^v+\alpha_n)^{c}\cap \mathcal{M}_n^v}\left\vert T_n^v(i)-T^v(i)\right\vert=o(1)$.

\end{theorem}

Because of the similar construction of all elements $T_{nl}(\cdot)$ of $T_n^v(\cdot)$, the consistency and inconsistency in different sets can be similarly investigated as the above for $T_n(\cdot)$. The estimation consistency of the locations and number of change points can still retain. We then do not give the results about $T_{nl}(\cdot)$ while directly stating the asymptotic results about the estimated locations and number of change points when the signal statistic $T_n^v(\cdot)$  is used.

\begin{theorem}\label{th2}

     Under the conditions in Proposition~\ref{com-con}, when  Conditions~6 and 7 in Supplementary Materials hold, $2\alpha_n\leq \alpha_n^{\star}$, and $\frac{n}{K\alpha_n^{\star}}\to \infty$, the results in Theorem~\ref{th1} hold true with respect to $T_n^v(\cdot)$.
  \end{theorem}

\subsection{A discussion on confidence interval construction}

The analysis for confidence interval construction and related issues in \cite{chen2021inference} can lead to similar results in our cases without structural mode, which are stated in Theorem \ref{cf} below. Before presenting the details, we give some notations used for the result.

Denote $$G_{i,l}=\frac{1}{2\alpha_n}\boldsymbol \Lambda^{-1}\boldsymbol \Sigma^{\frac{1}{2}}\left[I(1\leq i-l\leq \alpha_n)-I(1\leq l-i\leq \alpha_n)\right],$$ $\boldsymbol G_{i,\cdot}=(G_{i,1},G_{i,2},\cdots,G_{i,n})^{\top}$ and the jump size $\mathcal{M}_{\Delta}^{(k)}\colon =vec(\mathcal{M}^{(k+1)})-vec(\mathcal{M}^{(k)}),k=1,2,\cdots K.$

       Define $\mathcal{S}_{k}=\left\{1\leq j\leq p:\mathcal{M}_{\Delta,j}^{(k)}\neq 0\right\}$,
      where $\mathcal{M}_{\Delta,j}^{(k)}$ is the $j$th coordinate of $\mathcal{M}_{\Delta}^{(k)}$. Further, define $\tilde{\gamma}_k=\left((\boldsymbol \Lambda)^{-1}\mathcal{M}_{\Delta}^{(k)}\right)_{i\in \mathcal{S}_k},$ $\tilde{\tilde{\boldsymbol \Sigma}}_k=(\boldsymbol \Lambda^{-1}\boldsymbol \Sigma\boldsymbol \Lambda^{-1})_{i,j\in \mathcal{S}_k},$ $a_k=|\tilde{\gamma}_k|_2^2$ and $\zeta_k^2=\tilde{\gamma}_k^{\top}\tilde{\tilde{\boldsymbol \Sigma}}_k\tilde{\gamma}_k$. Note that
       $W(r)$ is a two-sided Brownian motion, that is, $W(r)=W_1(r)$ if $r>0$, and $W(r)=W_2(-r)$ if $r<0$, and $W_1$ and $W_2$ are two independent Brownian motions.
\begin{theorem}\label{cf} (The  confidence intervals of the locations of change points)
Under the conditions in Proposition \ref{con1} (ii),
      if $\min_{1\leq k\leq K}\min_{j\in \mathcal{S}_k}|(\boldsymbol \Lambda^{-1}\mathcal{M}_{\Delta}^{(k)})_j|\geq \omega^{+}\gg\sqrt{\frac{\log (np)}{\alpha_n}}+\frac{\alpha_n}{n}$ and for some constants $\underline{\lambda},\overline{\lambda}>0$, $\underline{\lambda}\leq \lambda_{max}(\boldsymbol \Lambda^{-1}\boldsymbol \Sigma\boldsymbol \Lambda^{-1})/\lambda_{min}(\boldsymbol \Lambda^{-1}\boldsymbol \Sigma\boldsymbol \Lambda^{-1})\leq \overline{\lambda},$
       then the confidence interval of $\frac{z_k^{\star}}{\alpha_n}$ at the  level $\alpha$ is  $\left(\frac{\hat{z}_k-\lfloor\hat{q}_{z,1-\frac{\alpha}{2}}\rfloor-1}{\alpha_n},\frac{\tilde{z}_k-\lfloor\hat{q}_{z,\frac{\alpha}{2}}\rfloor+1}{\alpha_n}\right)$, where $q_{z,1-\frac{\alpha}{2}}$ and $q_{z,\frac{\alpha}{2}}$ are $1-\frac{\alpha}{2}$ and $\frac{\alpha}{2}$th quantiles of the distribution of $\left(\frac{\zeta_k}{a_k}\right)^2\arg\max_{r}\left\{-2^{-1}|r|+W(r)\right\}$
       and $\hat{q}_{z,1-\frac{\alpha}{2}}$ and $\hat{q}_{z,\frac{\alpha}{2}}$ are the two estimators of the quantiles.
\end{theorem}
\begin{rema}
We discuss three issues in this remark.  First,  we can simulate quantiles by Monte Carlo method (\cite{chen2021inference}) or derive them from \cite{stryhn1996location} so that the confidence interval can be constructed.
   Second, note that the above theorem is based on the conditions in Proposition \ref{con1} (ii). Therefore, the above method may not be feasible under the conditions in  Proposition \ref{con1} (i) as in this case, $\eta_{ij}$  are not  i.i.d. and then the limiting distributions may not be derived. This deserves further study.
   Third,  when the mode-based distance is used to detect change points, the confidence intervals would be constructed in practice as follows. According to each change point, we can find a component $T_{nl}$ (though not unique) of $T_n^{v}$. Thus, we may use the similar results with respect to  the $l$th slice as those in Theorem~\ref{cf} to construct a confidence interval.  The details are omitted here.
\end{rema}

\section{Simulations}
To examine the performances of the proposed criterion, we conduct the simulations with several different model settings and compare our methods with existing competing methods, if any. For order-one tensor (vector), we compare with  the E-Divisive method (\cite{matteson2014nonparametric}), the change point detection through the Kolmogorov-Smirnov statistic(\cite{zhang2017pruning}), the change point detection via sparse binary segmentation(\cite{cho2015multiple}) and the Inspect method(\cite{wang2018high}). The following simulations abbreviate the four methods as ECP, KS, SBS, and Inspect. ECP and KS are implemented in the R package: ECP; SBS and Inspect are respectively implemented in the R packages: hdbinseg and InspectChangepoint and our method is SFD. We only consider our methods for higher-order tensors due to the lack of competitors.
In all simulation experiments, $n=1800$, $K=8$ and change point locations are respectively $200,400,600,800,1000,1200,1400,1600$. Each experiment is repeated $N=200$ times to compute the averages of the estimation $\hat{K}$, mean squared error(MSE), and the distribution of $\hat{K}-K$. Also, for the accuracy of the estimated locations, we compute the percentage of  ``correct'' estimations in the replications. Following \cite{cho2015multiple}, the estimated change points lie within the distance of $\lfloor \sqrt{n}/2\rfloor$  from the true change points is considered as the ``correct'' estimations. We calculate the proportion of all experiments in which at least four true change points were ``correctly" estimated, abbreviated as ``CP".

For the case with using the SF distance of MOSUM, we recommend that $\alpha_n=\lfloor \frac{2n^{3/4}}{9}\rfloor,$ $\epsilon_n=\frac{(\log n)^{0.55}}{\sqrt{\alpha_n}},$ $l_n(s)=s(\log n)^{1/2}\epsilon_n$,  and $c_n(i)=\frac{s_1\epsilon_n(\log n)^{0.55}}{I(i\in \mathcal{S})+\frac{1}{n}},$ where $s=2.5s_1$ and $s_1=1/50$. For the case with using the MSF distance of MOSUM, we recommend
$\alpha_n=\lfloor \frac{2n^{3/4}}{9}\rfloor,$ $\epsilon_n=\frac{(\log n)^{0.55}}{\sqrt{\alpha_n}},$ $l_n=s(\log n)^{1/2}\epsilon_n$, $c_{nl}(i)=\frac{s_1\epsilon_nn^{0.55}}{I(i\in \mathcal{S}_l)+\frac{1}{n}}$, $s=10s_1$, and $s_1=1/50$. Two basis algorithms are listed in Supplementary Materials.

\subsection{Order-one tensor}
Consider the model: $\boldsymbol x_i=\boldsymbol \mu_i+\boldsymbol \epsilon_i,$
where  $\boldsymbol\epsilon_i$  are independent error terms with mean zero, and  $p=50; 100; 500; 1,000; 2,000$. To save space, we only present the results for $p=50;100;2,000$, and the other results are presented in Supplementary Materials. 

{\itshape{Data generating mechanism:}} The data are segmented into nine parts and come from two distributions whose means are $\boldsymbol \mu_1$ and $\boldsymbol \mu_2$ respectively. $\boldsymbol \mu^{(i)}=\boldsymbol \mu_1$ for $i=1,3,5,7,9$ and $\boldsymbol \mu^{(i)}=\boldsymbol \mu_2$ for $i=2,4,6,8$. For space saving, we only present the result of dense case (See, Table~\ref{dense and strong}) 
and postpone the detailed results about sparse and mixed distribution in Supplementary Materials.
For dense case, we consider following cases:
  \begin{itemize}
    \item \textit{Strong signals.}  $\boldsymbol\epsilon_i \sim \mathcal{\boldsymbol N}(\boldsymbol 0,\boldsymbol I_{p\times p})$. All of the components of vectors $\boldsymbol \mu_1$ and $\boldsymbol \mu_2$ are equal to $1.4$ and $1$ respectively. The magnitude of the signal is then equal to $0.4$.
    \item \textit{Weak signals.} $\boldsymbol\epsilon_i \sim \mathcal{\boldsymbol N}(\boldsymbol 0,\boldsymbol I_{p\times p})$. The same setting as above except that all of the components of vectors $\boldsymbol \mu_1$ and $\boldsymbol \mu_2$ are equal to $1.2$ and $1$ respectively. The magnitude of the signal is  $0.2$.
  \end{itemize}
  We  also tried some other settings, such as the cases with dense (or sparse) and strong signals. Still, elements of the error terms are correlated, and the same settings as sparse and mixed distributions cases, except that all elements are independent. 
 As the comparisons with the other methods show similar observations to those from the above simulations, we then report the detailed results in Supplementary Materials.

 Our method is abbreviated as SFD. 
  The basic information from the simulation results is the following.
 First, SFD performs better with increasing dimensions, regardless of whether the covariance matrix is identity or non-identity. This dimensionality blessing for SFD deserves further study.  Second, for all the competitors, the performances are worse when the components of $\boldsymbol\epsilon_i$ are correlated, and thus, we consider the case with magnitudes being $1$. Third, ECP is a strong competitor, particularly when $p\le 100$ or $500$. However, SFD computes much faster. Fourth, detection in sparse cases is more difficult than in dense instances. Fifth, SFD  has, in comparison with the others, similar performance.

\begin{table}\tabcolsep=5pt
 \caption{\linespread{1.15} Dense case: distribution of $\hat K-K$.}\label{dense and strong}
 \small
 \bigskip
 \noindent\makebox[\textwidth][c]{
  \linespread{1.3}\selectfont
  \begin{tabular}{ccccccccccccccccccccc}
   \hline
   & & & & & & &$\hat K-K$ & &\smallskip \\  \cline{6-12}
   Scenarios & Procedures  &Means  &MSE  &CP &$\le-3$&$-2$&-1&0&1&2&$\ge 3$&  \smallskip \\ \hline
     \multicolumn{12}{c}{Signal=0.4; identity covariance matrix} \\
   (i)$p=50$
&SFD &$8.075$ & $0.175$ & $1$ & $0$ & $0$ & $10$ & $165$ & $25$ & $0$ & $0$ \\
&ECP & $8.080$ & $0.080$ & $1$ & $0$ & $0$ & $0$ & $184$ & $16$ & $0$ & $0$ \\
&KS & 5.015 & 10.105 & 0.395 & 136 & 47 & 14 & 3 & 0 & 0 & 0 \\
&SBS & 9.800 & 4.73 & 1 & 0 & 2 & 4 & 21 & 51 & 63 & 59 \\
&Inspect& 8.115 & 0.175 & 1 & 0 & 0 & 0 & 182 & 14 & 3 & 1 \\
(ii)$p=100$
&SFD &$8.075$ & $0.075$ & $1$ & $0$ & $0$ & $0$ & $185$ & $15$ & $0$ & $0$ \\
&ECP &$8.050$ & $0.060$ & $1$ & $0$ & $0$ & $0$ & $191$ & $8$ & $1$ & $0$ \\
&KS & 4.985 & 10.155 & 0.395 & 144 & 40 & 15 & 1 & 0 & 0 & 0 \\
&SBS & 9.985 & 5.115 & 1 & 0 & 0 & 2 & 15 & 47 & 71 & 65 \\
&Inspect  & 8.090 & 0.09 & 1 & 0 & 0 & 0 & 182 & 18 & 0 & 0 \\
(iii)$p=2000$
&SFD & 8.000 & 0 & 1 & 0 & 0 & 0 & 200 & 0 & 0 & 0 \\
&ECP & $8.045$ & $0.045$ & $1$ & $0$ & $0$ & $0$ & $191$ & $9$ & $0$ & $0$ \\
&KS &$4.985$ & $10.285$ & $0.300$ & $132$ & $55$ & $11$ & $2$ & $0$ & $0$ & $0$ \\
&SBS & $9.970$ & $5.140$ & $1$ & $0$ & $0$ & $2$ & $17$ & $44$ & $79$ & $58$ \\
&Inspect &$8.580$ & $1.150$ & $1$ & $0$ & $0$ & $0$ & $122$ & $54$ & $14$ & $10$ \\
\hline \\[-1.8ex]
 \multicolumn{12}{c}{Signal=0.2; identity covariance matrix} \\
   (i)$p=50$
&SFD &$7.100$ & $1.800$ & $0.925$ & $15$ & $40$ & $60$ & $80$ & $5$ & $0$ & $0$ \\
&ECP &$8.075$ & $0.105$ & $1$ & $0$ & $0$ & $0$ & $187$ & $12$ & $0$ & $1$ \\
&KS &$4.835$ & $11.355$ & $0.040$ & $147$ & $38$ & $12$ & $3$ & $0$ & $0$ & $0$ \\
&SBS &$9.960$ & $5.310$ & $0.985$ & $0$ & $1$ & $4$ & $18$ & $41$ & $71$ & $65$ \\
&Inspect &$5.515$ & $17.455$ & $0.635$ & $79$ & $24$ & $6$ & $48$ & $34$ & $8$ & $1$ \\
(ii)$p=100$
& SFD &$7.950$ & $0.400$ & $0.975$ & $0$ & $0$ & $35$ & $150$ & $5$ & $10$ & $0$ \\
&ECP & $8.080$ & 0.09 & 1& 0 & 0 & 0 & 185 & 14 & 1 & 0 \\
&KS & $4.850$ & $10.970$ & $0.020$ & $152$ & $39$ & $8$ & $1$ & $0$ & $0$ & $0$ \\
&SBS & $9.875$ & $4.845$ & $1$ & $0$ & $0$ & $1$ & $19$ & $61$ & $61$ & $58$ \\
&Inspect &$7.790$ & $4.880$ & $0.910$ & $20$ & $9$ & $1$ & $105$ & $39$ & $21$ & $5$ \\
(iii)$p=2000$
&SFD &$7.875$ & $0.125$ & $1$ & $0$ & $0$ & $25$ & $175$ & $0$ & $0$ & $0$ \\
&ECP & $8.065$ & $0.075$ & $1$ & $0$ & $0$ & $0$ & $188$ & $11$ & $1$ & $0$ \\
&KS &$4.760$ & $11.590$ & $0.035$ & $158$ & $29$ & $13$ & $0$ & $0$ & $0$ & $0$ \\
&SBS & $10.140$ & $5.740$ & $1$ & $0$ & $0$ & $0$ & $13$ & $41$ & $72$ & $74$ \\
&Inspect & $9.020$ & $3.210$ & $1$ & $0$ & $0$ & $0$ & $99$ & $51$ & $24$ & $26$ \\
\hline
\end{tabular}}
\end{table}
\clearpage
\subsection{Higher order tensor}
%
%
%
%
%
%
%
%
%
 As the literature lacks change point detection methods for general order-two and -three tensors, we only report the simulation results of the two proposed criteria in this paper.
Consider the tensor model as \begin{equation}\label{matrix}
   \boldsymbol{{X}_i}=\boldsymbol{\Sigma_i}+\boldsymbol{E_i},i=1,2,\cdots,n.
\end{equation}
When $\boldsymbol E_i$ is a random matrix whose elements $E_{i, jk}$ are with means zero, the model is order-two. Consider $p_2=kp_1$ for $k=1,16.$
Similar observations are made on the simulations with order-three tensors, which are then put in Supplementary Materials.


{\itshape{Data generating mechanism:}} The data are segmented into nine parts coming from two distributions whose means are $ \boldsymbol{\Sigma_1}$ and $\boldsymbol{\Sigma_2}$ respectively. $\boldsymbol{\Sigma^{(i)}}=\boldsymbol{\Sigma_1}$ for $i=1,3,5,7,9$ and $\boldsymbol{\Sigma^{(i)}}=\boldsymbol{\Sigma_2}$ for $i=2,4,6,8$.

In the simulations, each row of $\boldsymbol{E_i}$ follows $\mathcal{N}(0,\boldsymbol{\tilde{\Sigma})}$ and rows are independent of each other.
When $p_2=p_1$, all elements in $\boldsymbol{\Sigma_1}$ and $\boldsymbol{\Sigma_2}$ are equal to $1.4$ and $1$, respectively. Here $p_1=10,30,50$. The magnitude of the signal is $0.4$.  When $p_2=16p_1$, For $\boldsymbol{\Sigma_1}=(\sigma_{ij})$, $\sigma_{ij}=0.8^{|i-j|}$ if $j\leq i$; otherwise, $\sigma_{ij}=1$. All elements in $\boldsymbol{\Sigma_2}$ are equal to $1$. Here $p_1=10,12$. For space-saving, the cases with all independent elements and order-three tensors are shown in Supplementary Materials.
\begin{table}[h]\tabcolsep=5pt
 \caption{\linespread{1.15} Order-two tensor(row correlation): The  distribution of $\hat K-K$.}\label{compare1 corr}
 \small
 \bigskip
 \noindent\makebox[\textwidth][c]{
  \linespread{1.3}\selectfont
  \begin{tabular}{ccccccccccccccccccccc}
   \hline
   & & & & & & &$\hat K-K$ & &\smallskip \\  \cline{6-12}
   Scenario &Procedure &Mean  &MSE  & CP &$\le-3$&$-2$&-1&0&1&2&$\ge 3$&  \smallskip \\ \hline
     \multicolumn{12}{c}{Symmetric cases: $p_1=p_2$} \\
$p_1=10$ &MSFD       & $3.205$ & $24.805$ & $0.230$ & $188$ & $9$ & $3$ & $0$ & $0$ & $0$ & $0$ \\
         &SFD &$8$ & $0.430$ & $1$ & $0$ & $3$ & $32$ & $129$ & $34$ & $2$ & $0$ \\
$p_1=30$ &MSFD        &$8.020$ & $0.380$ & $1$ & $0$ & $3$ & $26$ & $136$ & $34$ & $1$ & $0$ \\
       &SFD &$8.005$ & $0.005$ & $1$ & $0$ & $0$ & $0$ & $199$ & $1$ & $0$ & $0$ \\
$p_1=50$ &MSFD        &$8.295$ & $0.515$ & $1$ & $0$ & $0$ & $7$ & $140$ & $42$ & $9$ & $2$ \\
         &SFD &$8$ & $0$ & $1$ & $0$ & $0$ & $0$ & $200$ & $0$ & $0$ & $0$ \\
 \hline
 \multicolumn{12}{c}{Asymmetric cases: $p_2=16p_1$} \\
$p_1=10$ &MSFD         &$7.365$ & $1.825$ & $0.765$ & $10$ & $34$ & $67$ & $57$ & $25$ & $6$ & $1$ \\
         &SFD & $5.595$ & $7.485$ & $0.920$ & $81$ & $69$ & $41$ & $9$ & $0$ & $0$ & $0$ \\
$p_1=12$ &MSFD        &$7.960$ & $0.900$ & $0.925$ & $0$ & $14$ & $39$ & $100$ & $36$ & $10$ & $1$ \\
       &SFD  &$7.010$ & $1.930$ & $0.995$ & $15$ & $33$ & $82$ & $70$ & $0$ & $0$ & $0$ \\
\hline
\end{tabular}}
\end{table}

The results reported in Table~\ref{compare1 corr} suggest that the symmetric case favors the SFD-based criterion, whereas the asymmetric case with a much large mode favors the mode-based SFD method. This is also the case when the elements of error matrices are independent. See Supplementary Materials.

\section{Application}
This section gives three real data examples concerned with order-two and order three tensors. We also put the analysis for the order-three tensor data set of video to Supplementary Materials.

\subsection{Enron email dynamic network}
 The CALO project contains data from about $150$ users and their email exchanges. The processed data were used in \cite{priebe2005scan}, based on $184$ unique addresses over 189 weeks from 1998 to 2002. The data are available in
 \url{https://www.cis.jhu.edu/~parky/Enron/}. We extract the information about the time for sending and receiving emails from each address during the $189$ weeks. We aim to check whether the mail exchange patterns change during this period, reflecting changes in the relationship among these addresses. Each address is regarded as one subject, called the node in networks. The dynamic networks' adjacency matrix $\boldsymbol A_i,i=1,2,\cdots,188$ reflects the links between any two subjects. If there is an email exchange between subjects $i$ and $j$, the $(i,j)$ element of $\boldsymbol A_i$ is $1$; otherwise, it is zero.

 As this tensor is order-two with $p_1= p_2$, we first use the SFD-based statistic to detect change points.   Three change points are identified at the locations  $39$, $94$, and $141$. Thus, the email exchange patterns may change around the July 1999, August 2000 and July 2001. As the sample size is relatively small compared with the size of the tensor, we also use the MSFD-based statistic that identifies the locations  $54$,   $111$, and $149$, which correspond to November 1999, December 2000 and September 2001. 

 \cite{priebe2005scan} made anomaly detection and identified three weeks $58$th, $94$th and $146$th weeks through their scan statistics with order $0$ and order $1$. These are similar to ours.
 \cite{enron} recorded Enron Chronology and stated the following facts: Enron purchased 5.1 percent of the long-distance gas pipeline company in July 1999 ($54$th week), sham energy deal occured in December 1999 ($58$th week),   large-scale insider selling was underway in August 2000 ($94$th week), Enron bought back subsidiary's publicly held stock in December 2000 ($111$th week), Enron's stock closed at $46.66$ in July 2001 ($141$th week),  Enron CEO made a resignation announcement in September 2001 ($146$th week) and it was announced that Enron  restructured itself in September 2001 ($149$th week).
Thus, our detection could reflect those changes that happened in Enron.

\subsection{Seismic Data}
\begin{figure}[h]
  \centering
  \includegraphics[width=0.6\textwidth]{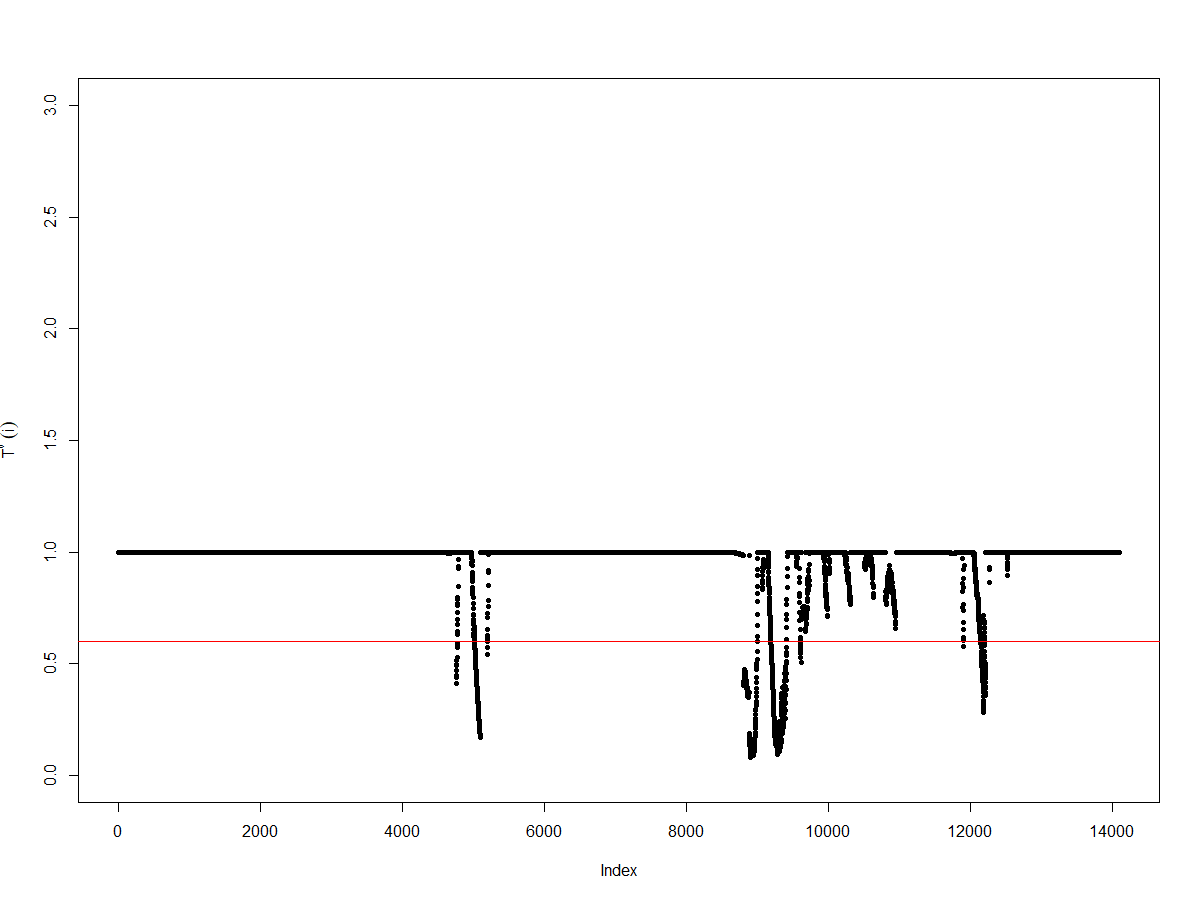}
  \caption{The plot presents the curve of $T_n^v(\cdot)$ in seismic data 
  }\label{seismic}
\end{figure}

This dataset is retrieved from \url{http://service.ncedc.org/fdsnws/dataselect/1/.} It has been analyzed in \cite{xie2019asynchronous} and \cite{chen2022high}. This dataset contains ground motion sensor measurables recorded from 39 sensors explained that it is important for fusion center to declare an alarm whenever a change point occurs in any sensor.  \cite{chen2022high} removed the linear trend and temporal dependence of the data and the processed data are included in R Package ocd. The processed data set contains 14998 observations corresponding to a time from 2.00-2.16am on December 23, 2004 and each observation is a 39-dimensional vector. We have also known a fact that there is an earthquake measured at duration magnitude 1.47Md hit near Atascadero, California at 02:09:54.01. The structural mode consists of 39 sensors, and then we apply MSFD to obtain four change points at $364.352, 606.08, 637.056$ and $818.176$  corresponding to the time points $02:06:04.352, 02:10:06.08, 02:10:37.056$ and $ 02:13:38.176$, respectively. Figure \ref{seismic} presents the curve of $T_n^v(\cdot)$.  

\section{Conclusion}

When a structural mode needs to be separately treated,  we in this paper propose a general  mode-based signal-screening Frobenius distance (MSFD) of MOSUM.  so that we can handle dense and sparse scenarios. When all  modes can be equally treated, this distance is reduced to be a signal-screening Frobenius distance (SFD) of MOSUM. The defined criterion is based on ratios of MSFD's with adaptive-to-signal ridge functions to enhance the detection capacity.
The method has several advantages: estimation consistency, robustness against distribution and sparsity structure, computational simplicity, and visualization.  Its limitations are mainly in two aspects. We require that the distance between any two changes cannot be too small and the performances of the criterion are relatively sensitive to the choices of $s_1$ in the ridge and $s$ in the threshold in the distance. These deserve further study.

\begin{center}
{\large\bf SUPPLEMENTARY MATERIAL}
\end{center}

\begin{description}
	
	\item {\bf Supplemenraty of Multiple change point detection of tensor} Conditions, proofs of theoretical results and some additional numerical studies . (.pdf file)

\end{description}
	
	

\bibliographystyle{Chicago}


\bibliography{reference}

\end{document}